\documentclass[a4paper,11pt]{article}
\usepackage[utf8]{inputenc}
\usepackage{graphicx}
\usepackage{xfrac}
\usepackage{amsmath,amsfonts,amssymb}
\usepackage{subcaption}
\usepackage{color}
\usepackage{slashed}
\usepackage{cite}
\usepackage{csquotes}
\usepackage{tabularx}
\usepackage{authblk}
\usepackage[normalem]{ulem}
\usepackage[colorlinks=true, urlcolor=blue, linkcolor=blue, citecolor=blue]{hyperref}
\usepackage[toc,page]{appendix}


\newcommand{\KaTie}{Ka\hspace{-0.2ex}Tie}

\title{%
\hspace{\fill}{\normalsize IFJPAN-IV-2021-13}\\[4ex]
\bf $D$ and $B$-meson production using $k_t$-factorization calculations in a variable-flavor-number scheme}
\author[1]{B. Guiot\thanks{benjamin.guiot@usm.cl}}
\author[2]{A. van Hameren\thanks{hameren@ifj.edu.pl}}
\affil[1]{Departamento de F\'isica, Universidad T\'ecnica Federico Santa Mar\'ia; Casilla 110-V, Valparaiso, Chile}
\affil[2]{Institute of Nuclear Physics, Polish Academy of Sciences, Radzikowskiego 152, 31-342 Krak\'ow, Poland }

\date{}

\begin{document}

\maketitle

\begin{abstract}
Within the framework of $k_t$-factorization, we compute the differential cross section for the production of $B$ and $D$ mesons, using a general-mass variable-flavor-number scheme. Our calculations include all relevant $2\to 2$ processes. We explain how to include the $2\to 1$ process in our calculations, but argue this is not (numerically) relevant at moderate transverse momentum due to its cancellation with the subtraction term. We apply this formalism to $pp$ collisions and compare our results with ALICE and LHCb data at central and forward rapidity. 
\end{abstract}
\newpage

\tableofcontents

\section{Introduction}
Heavy flavors, which play a particular role in perturbative Quantum ChromoDynamics (pQCD), have been extensively studied. In the beginning, the observable was the total cross section, plotted as a function of the center-of-mass energy $\sqrt{s}$. Later, with higher energies and statistics available, the differential cross section has been measured, for instance, at the Tevatron and Large Hadron Collider (LHC). We note $y$ as the rapidity of the detected heavy flavor and $p_t$ as its transverse momentum. The theoretical description of this observable, requiring a more careful treatment compared to the total cross section, has known some troubles \cite{Cacc}. In the framework of collinear factorization, it is known that general-mass variable-flavor-number schemes (GM-VFNS) are more efficient than fixed-flavor-number schemes (FFNS) to describe the differential cross section at large $p_t$. By more efficient, we mean that, at a given order, the former gives better results than the latter, which is particularly true at leading order (LO). This better efficiency is partially explained by the fact that the GM-VFNS resums to all orders some large logarithms $\ln p_t^2/m_Q^2$, thanks to the heavy-quark distribution function.\\

Heavy-quark production has also been addressed by the first works on $k_t$ factorization \cite{CaCiHa,ktFac,ktFac2,semihard2}. During the last two decades, heavy-flavor data have been compared to $k_t$ factorization predictions by several groups, see, for instance, \cite{LiSaZo,jun02,JuKrLi,MaSz,ShShSu,MaSz2}. However, as pointed out in \cite{gui2,gui3}, some of the available calculations are not performed consistently with respect to the choice of the scheme. One reason for this is simply the lack of appropriate unintegrated parton distributions (uPDFs). Another reason is the complications coming from the use of off-shell matrix elements and the difficulty to compute the cross section at higher orders. The main goal of the present work is to provide for the first time $k_t$-factorization calculations for $D$ and $B$-meson production, using $2\to 2$ processes evaluated in a GM-VFNS. The $2\to 1$ process with an initial off-shell charm has already been addressed in \cite{Kniehl:2008qb}.\\

In Sec.~\ref{seckt}, we have a general discussion on the schemes used in calculations. Then, we present our GM-VFNS uPDFs in Sec.~\ref{secwmr} and the event generator \KaTie\ used for the evaluation of the off-shell cross section in Sec.~\ref{seckatie}. In Sec.~\ref{secres}, we compare our result to ALICE and LHCb data. Our calculations include all relevant $2\to 2$ processes, which in $k_t$-factorization are next-to-leading order (NLO) contributions. In Sec.~\ref{seccompth} we perform a comparison of our results for $D$ and $B$ mesons with other $k_t$-factorization calculations. In Sec.~\ref{sec12}, we discuss the implementation of the LO, $2\to 1$, process and argue that it can be ignored at moderate transverse momentum, defined by $·\alpha_s\ln p_t/m_Q \sim \mathcal{O}(1)$, due to its cancellation with the subtraction term. Finally, we give our conclusion in Sec.~\ref{seccon}.

\section{Unintegrated PDFs: discussion on the scheme \label{seckt}}
In collinear factorization, the cross section for heavy-quark production in hadron-hadron collisions is given schematically by
\begin{equation}
    \sigma=\sum_{i,j}f_{i/h}\otimes f_{j/h} \otimes \hat{\sigma}(ij\to Q +X),\label{colfac}
\end{equation}
where $f_{k/h}$ is the collinear parton distribution for the parton $k$ in the hadron $h$, and $\hat{\sigma}$ is the partonic cross section. Collinear PDFs are extracted by comparison of Eq.~(\ref{colfac}) with data. It is then clear that the output $f_{k/h}$ depends on the input $\hat{\sigma}$. A larger partonic cross section requires smaller PDFs, and the choice made for $\hat{\sigma}$ defines the scheme. Once the PDFs have been extracted, they can be used to make predictions. However, the new partonic cross section should be computed with the scheme used in the extraction of the PDFs.\\

The situation is similar in $k_t$ factorization. The cross section reads
\begin{multline}
 \frac{d\sigma}{dx_1dx_1d^2p_t}(s,x_1,x_2,p_t^2)=\sum_{i,j}\int^{k_{t,\text{max}}^2}_0 d^2k_{1t}d^2k_{2t} F_{i/h}(x_1,k_{1t}^2;\mu^2)\\
 \times F_{j/h}(x_2,k_{2t}^2;\mu^2)\hat{\sigma}(x_1x_2s,k_{1t}^2,k_{2t}^2,p_t^2;\mu^2), \label{ktfac}
\end{multline}
where the uPDFs, $F_{k/h}(x,k_{t}^2;\mu^2)$, depend on $x$, the fraction of the hadron longitudinal momentum carried by the parton, $k_t$, the initial parton transverse momentum, and $\mu$, the factorization scale. $\hat{\sigma}$ is the off-shell cross section. The uPDFs are generally not extracted from data\footnote{An exception is the PB uPDFs \cite{PB}.} but built from the collinear PDFs by inverting the relation
\begin{equation}
    f_{k/h}(x,\mu^2)=\int^{\mu^2}_0 F_{k/h}(x,k_t^2;\mu^2)dk_t^2. \label{undens}
\end{equation}
Note that different versions of this relation can be found in the literature. It is clear that the scheme of the uPDFs built from Eq.~(\ref{undens}) should be identified to the scheme of the collinear PDFs appearing in this equation. Everything we said on the scheme is also true for the order of calculation. As a consequence, the different sets of uPDFs available on the TMDlib \cite{tmdlib,tmdlib2} cannot be compared by simply using them with the same cross section, as they have been obtained in different schemes and at different orders. For our present study, we use uPDFs and off-shell cross sections obtained in a GM-VFNS at order $\mathcal{O}(\alpha_s^2)$.\\

The numerical consequence of using a cross section at an order/scheme different from that of uPDFs depends on the case. Mixing the VFNS and FFNS could lead to the wrong estimation of the cross section for charm production by a factor of 4 \cite{gui2,gui3}. Indeed in the VFNS, the cross section obtained at order $\mathcal{O}(\alpha_s^2)$ reads
\begin{equation}
    \sigma(\text{charm})=\sum_{i,j}f_i^{\text{VFNS, (2)}}\otimes f_j^{\text{VFNS, (2)}}\otimes\hat{\sigma}^{\text{(2)}}(\text{FEP +  FCP}), \label{vfnscharm}
\end{equation}
where FEP stands for flavor excitation processes, e.g., $cg\to cg$, and FCP for flavor creation processes, e.g., $gg\to c\Bar{c}$.\footnote{The number two in parenthesis indicates that we work at the order $\mathcal{O}(\alpha_s^2)$.} On the opposite, using the FFNS, we have
\begin{equation}
    \sigma(\text{charm})=\sum_{k,l}f_k^{\text{FFNS, (2)}}\otimes f_l^{\text{FFNS, (2)}}\otimes\hat{\sigma}^{\text{(2)}}(\text{FCP}). \label{ffnscharm}
\end{equation}
We changed the subscript to $k$ and $l$ to point out that the sums in Eqs. (\ref{vfnscharm}) and (\ref{ffnscharm}) are not identical. We know that $\hat{\sigma}^{\text{(2)}}(\text{FEP +  FCP})\sim 4 \hat{\sigma}^{\text{(2)}}(\text{FCP})$, due to the large contribution of the flavor excitation process $cg\to cg$. Then, we deduce that
\begin{equation}
    \sum_{i,j}f_i^{\text{VFNS, (2)}}\otimes f_j^{\text{VFNS, (2)}}\sim \frac{1}{4 }\sum_{k,l}f_k^{\text{FFNS, (2)}}\otimes f_l^{\text{FFNS, (2)}}.
\end{equation}
Consequently, using the VFNS PDFs with $\hat{\sigma}(\text{FCP})$ implies an underestimation of the cross section by a factor of 4. The (numerical) situation could improve at higher orders if $\hat{\sigma}(\text{FCP})$ and $\hat{\sigma}(\text{FEP + FCP})$ are numerically closer, but mixing different schemes is always inconsistent and dangerous.\\

Strictly speaking, calculations performed in \cite{gui2} are not completely consistent since the NLO PB uPDFs \cite{PB}, being, in fact, next-to-next-to-leading order (NNLO) in $k_t$-factorization, have been combined with a NLO cross section ($\mathcal{O}(\alpha_s^2)$). However, the numerical deviation from the calculations performed in the present paper is found to be small. 

The main goal of the present work is to provide consistent GM-VFNS $k_t$-factorization calculations. Consequently, we cannot use the PB uPDFs, and we build our own NLO (in $k_t$-factorization) VFNS uPDFs in the next section.

\section{The Watt-Martin-Ryskin unintegrated PDFs \label{secwmr}}
Accordingly to Eq.~(\ref{undens}), VFNS uPDFs of order $\mathcal{O}(\alpha_s^2)$ can be built from collinear PDF extracted in this scheme and at this order. In the Watt-Martin-Ryskin (WMR) approach \cite{wmr}, the uPDF for a parton $a$ reads
\begin{equation}
F_a(x,k_t^2;\mu^2)=\frac{1}{k_t^2}T_a(\mu,k_t)\sum_{a'}\int_x^{1-\Delta}\frac{dz}{z}\frac{\alpha_s(k_t)}{2\pi}P_{aa'}(z)f_{a'}\left(\frac{x}{z},k_t\right).\label{def2}
\end{equation}
where $P_{aa'}(z)$ are the usual unregulated splitting functions, except for $P_{gg}$ given by
\begin{equation}
    P_{gg}=2C_A\left[\frac{z}{1-z}+\frac{1-z}{z}+z(1-z)  \right],
\end{equation}
and $T_a(\mu,k_t)$ is the Sudakov factor defined by
\begin{equation}
T_a(\mu,k_t)=\exp \left\lbrace -\int_{k_t^2}^{\mu^2}\frac{dp_t^2}{p_t^2}\sum_{a'}\int_0^{1-\Delta(p_t)}dz\,z\frac{\alpha_s(p_t)}{2\pi}P_{a'a}(z)\right\rbrace. \label{suda}
\end{equation}
A possible choice for $\Delta$ is the strong ordering cutoff
\begin{equation}
    \Delta(q)=\frac{q}{\mu},
\end{equation}
with $\mu\sim \sqrt{p_t^2+m^2}$ the factorization scale. With this cutoff, 
\begin{equation}
F_a(x,k_t^2;\mu^2)=0 \quad \text{ if}\quad k_t>\mu.  \label{condSO}  
\end{equation}
However, the strong ordering cutoff does not stem from fundamental considerations, and the obtained uPDFs do not fulfill relation (\ref{undens}) with good precision. This is because $\Delta$ is too large and cuts half of the phase space when $k_t=0.5\mu$. Consequently, we keep the condition (\ref{condSO}) and use $\Delta/d(x)$, with $d(x)>1$ chosen such that the relation (\ref{undens}) is satisfied to a good approximation.\\

\begin{figure}[!h]
\centering
 \includegraphics[width=30pc]{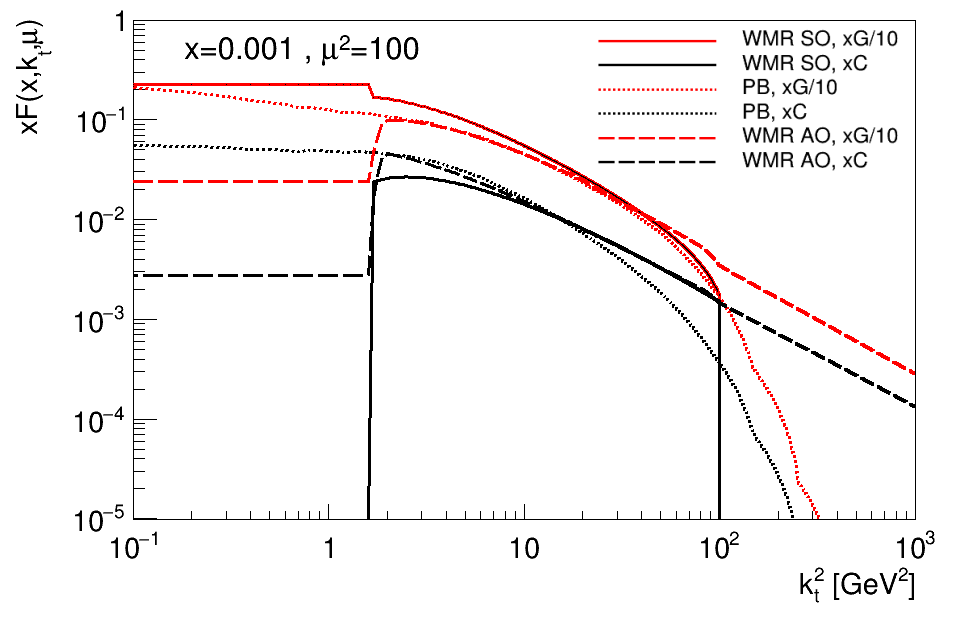}
\caption{Comparison of our SO WMR uPDFs (full lines) with the PB (dotted lines) and AO WMR (dashed lines) uPDFs. \label{uncomp}}
\end{figure}

\begin{figure}[!h]
\centering
 \includegraphics[width=28pc]{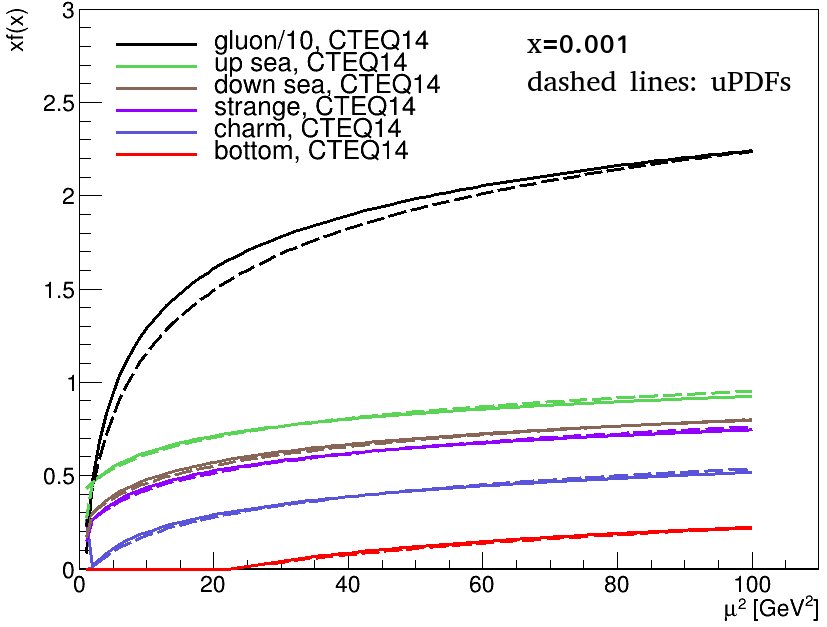}
\caption{Comparison of CT14 LO parton distributions (full lines) at $x=0.001$ with the integrated uPDFs (dashed lines), see Eq.~(\ref{undens}). \label{unpdf1}}
\end{figure}

\begin{figure}[!h]
\centering
 \includegraphics[width=27.5pc]{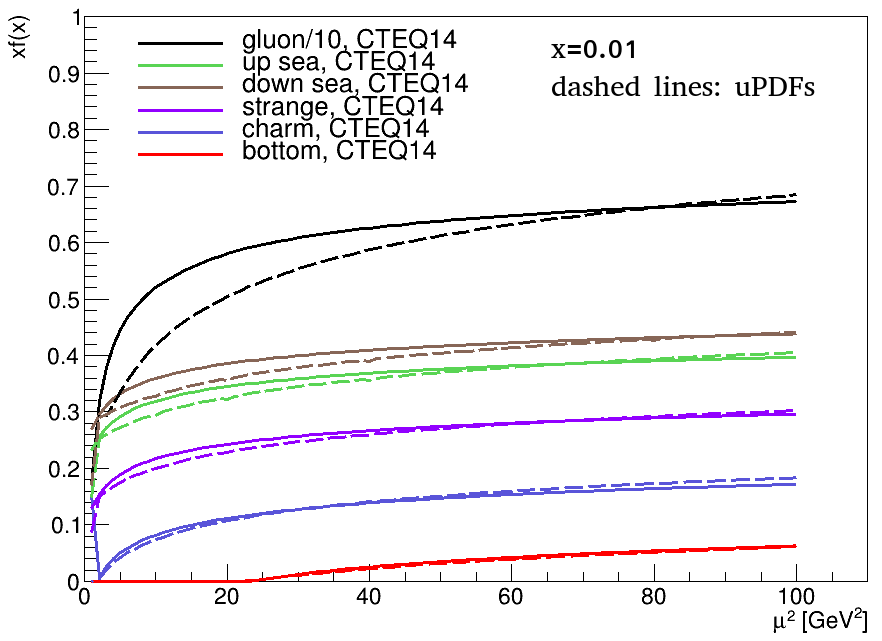}
\caption{Same as Fig. \ref{unpdf1} for $x=0.01$.\label{unpdf2}}
\end{figure}

We built our uPDFs using the WMR formalism and the leading order CT14 PDFs \cite{ct14}. These collinear PDFs have been extracted using the $\text{ACOT}_\chi$ scheme \cite{acot1,acot2,acot3}. In Fig.~\ref{uncomp}, we compare our unintegrated charm and gluon distributions with the PB and angular-ordering WMR results, obtained from the TMDlib. The main differences are encountered at small $k_t$ and at $k_t^2>\mu^2$. In particular, we observe the typical slow falloff of the AO WMR distribution, leading to an overestimation of the heavy-quark cross section \cite{gui2}.
In Figs.~\ref{unpdf1} and \ref{unpdf2}, we plot Eq. (\ref{undens}) as a function of $\mu^2$ for different values of $x$. We observe a global good agreement between the integrated distributions and the collinear PDFs, in particular, at small $x$.

\section{The \KaTie\ event generator \label{seckatie}}
The calculations in the following have been performed with the help of the parton-level event generator \KaTie~\cite{vanHameren:2016kkz}.
It can generate events for which the partonic initial-state momenta are spacelike and have nonvanishing transverse components.
The necessary uPDFs can be provided by TMDlib or as grid files in text format.
In the latter case, \KaTie\ takes care of the interpolation.
\KaTie\ operates at tree level and can deal with any processes within the standard model.
It generates events employing importance sampling and numerically evaluates the matrix element using helicity amplitudes.
The events are stored in event files which can be chosen to be in the LHEF format~\cite{Alwall:2006yp}.
It is, however, also possible to make histograms directly using the provided tools. 

The initial-state and final-state momenta are generated satisfying exact kinematics.
Within $k_t$ factorization, this means that the initial-state momenta $k_{i}^\mu=x_{i}P_{i}^\mu + k_{i\,t}^\mu$ are spacelike, with a longitudinal component along the lightlike momentum $P_i^\mu$ of either colliding hadron, plus transverse components $k_{i\,t}^\mu$.
The matrix elements are constructed as described in~\cite{vanHameren:2012if,vanHameren:2013csa}, as summed squares of helicity amplitudes.
The essence of the method is that a spacelike external parton is  represented as a pair of auxiliary lightlike partons satisfying eikonal Feynman rules, leading to manifestly gauge invariant amplitudes with exact kinematics.

In this paper, we consider parton-level processes that, within collinear factorization, would involve massive initial-state quarks with timelike momenta.
The same construction of the amplitudes with spacelike initial-state momenta leads also for these processes to manifestly gauge invariant matrix elements.
The only restriction is that the transverse momentum of the initial-state parton must not be much smaller than the mass.
While gauge invariance is guaranteed, the correct on-shell limit cannot be reached for timelike momenta by naively taking very small transverse momentum.
Note that the kinematics of the final-state massive quark is always exact, with its momentum timelike $p^2=m_Q^2$.

As mentioned earlier, \KaTie\ operates at tree level.
The tree-level matrix elements contain singularities as functions of the external momenta when these become soft or collinear with each other.
In particular, the matrix elements behave singularly if a final-state gluon momentum becomes collinear with the momentum of an initial-state hadron.
The latter also happens in $k_t$ factorization.
We will consider partonic processes with a final-state gluon, which suffer from such a collinear singularity.
In usual tree-level calculations, these singularities are avoided by phase space cuts that define the jet observables, for example, the demand of a minimal transverse momentum.
We also use a small minimum on the transverse momentum to avoid the singularity.
We see that the cross section depends only very mildly on this phase space cut if we vary the minimum between $0.5$ and $2.0$ GeV, the latter being of the order of the mass of the quarks we are considering.

\section{Comparison with data \label{secres}}
\subsection{$D$-meson production}
Working with a GM-VFN scheme, the following processes are included
\begin{align}
    gg&\to Q\bar{Q} \quad q\bar{q}\to Q\bar{Q}\label{fcre}\\
    gQ&\to gQ \quad qQ\to qQ \quad \bar{q}Q\to \bar{q}Q,\label{fexc}
\end{align}
where $Q$ represents the heavy quark and $q$ a light quark. The first and second lines correspond to flavor-creation and flavor-excitation processes, respectively. For consistency, we choose the charm mass equal to the one used for the CT14 PDFs, e.g., $m_c=1.3$ GeV. For the fragmentation of a charm quark into a $D$ meson, we use the Peterson model of fragmentation function \cite{pescsc} with $\epsilon_c=0.05$, and the fragmentation fractions given in table \ref{fra2}.
\begin{table}[ht]
\centering
\begin{tabular}{ | m{2.2cm} | m{2.2cm}| m{2.2cm} | m{2.2cm} |} 
\hline
 $f(c\to D^0)$ & $f(c\to D^+)$ & $f(c\to D^{*^+})$ & $f(c\to D_s^+)$\\
 \hline
 0.588  & 0.234  & 0.234 & 0.116 \\
\hline
\end{tabular}
\caption{Charm to $D$-meson fragmentation fractions.}
\label{fra2}
\end{table}
The three first values are those used by FONLL \cite{fonll}.\\

\begin{figure}[!h]
\centering
 \includegraphics[width=20pc]{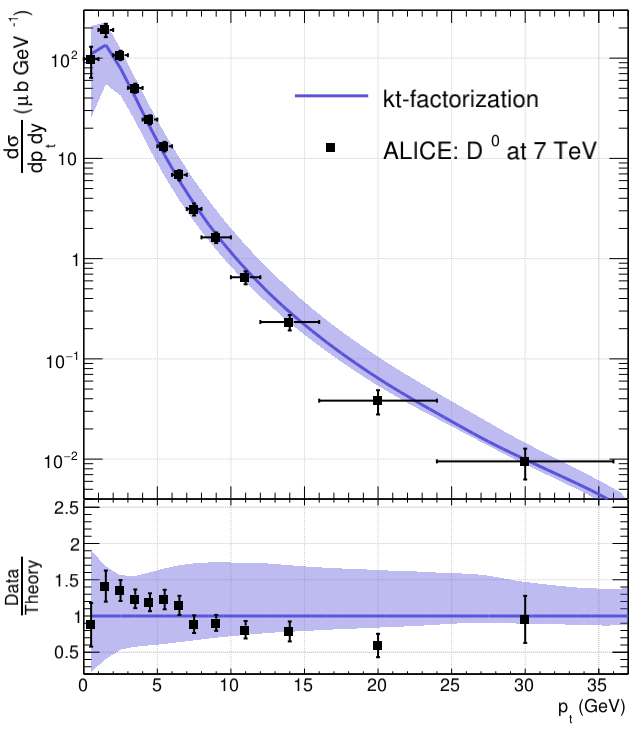}
\caption{Leading order $k_t$-factorization calculations, obtained in a GM-VFNS, compared to ALICE data \cite{al7D2017} for $D^0$ mesons. The line shows the central value of our calculations and the band corresponds to the factorization scale uncertainty.\label{d07}}
\end{figure}

The result for the pt distribution of $D^0$ mesons is compared to ALICE data \cite{al7D2017} in Fig.~\ref{d07}.
The error band corresponds to the factorization scale uncertainty, evaluated as usual by the variation of a factor of $\sqrt{2}$ above and below the central value, chosen to be
\begin{equation}
    \mu=\frac{1}{2}(m_{t,1}+m_{t,2}),\label{scalekt}
\end{equation}
where $m_{t,i}=\sqrt{p_{t,i}^2+m_c^2}$. The subscripts $1$ and $2$ label outgoing partons. One of these partons is a charm, since processes such as $gg\to gg$ with a final gluon fragmenting into a $D$ meson have not been considered\footnote{An example of scale-dependent fragmentation functions in $k_t$ factorization, including the contribution $g\to D$, can be found in Refs. \cite{KaNeSaC,KaNeSaB}.}. Note that we use $m_{t,i}$ even if $p_{t,i}$ corresponds to a gluon or light quark transverse momentum. In collinear factorization, the usual choice for the factorization scale is 
\begin{equation}
\mu=m_t,\label{scalecol}
\end{equation}
with $m_t$ the charm transverse mass. In the limit where the transverse momentum of the initial partons goes to zero, $k_{1t}, \, k_{2t}\to 0$, Eqs. (\ref{scalekt}) and (\ref{scalecol}) coincide.
\begin{figure}[!h]
\centering
 \includegraphics[width=20pc]{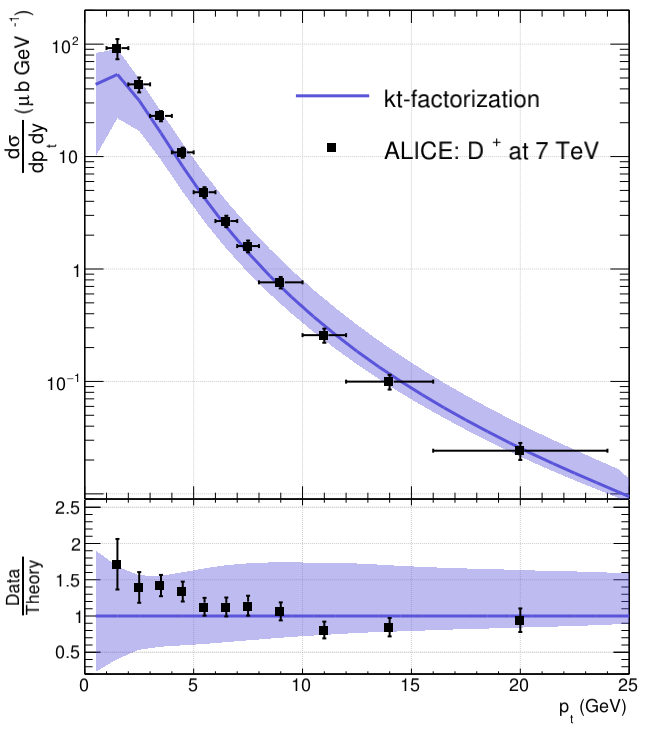}
\caption{Same as Fig. \ref{d07} for $D^+$ mesons. \label{dp7}}
\end{figure}
In Fig.~\ref{dp7}, we compare our calculations for the $p_t$ distribution of $D^+$ mesons with ALICE data. We observe a good agreement on the full $p_t$ range, and the central value alone provides a good description, except for the first bin of Fig. \ref{dp7}. A Comparison with figure 5 of Ref. \cite{al7D2017} shows that the underestimation of $D$-meson data at small transverse momentum by theoretical calculations is usual. We will see in the next section that this is not the case for $B$ mesons. Based on \cite{al7D2017}, where theoretical calculations are compared
with the measured cross section, we observe that our work represents a significant improvement of $k_t$-factorization calculations. We believe this is directly related to the consistent use of a GM-VFNS. To reach a similar result in an FFNS, it is probably necessary to include higher orders. We compare our results to other $k_t$-factorization calculations in more detail in section \ref{seccompth}. Note also that our central values are slightly better in comparison with FONLL calculations.  

\begin{figure}[!h]
\centering
 \includegraphics[width=22pc]{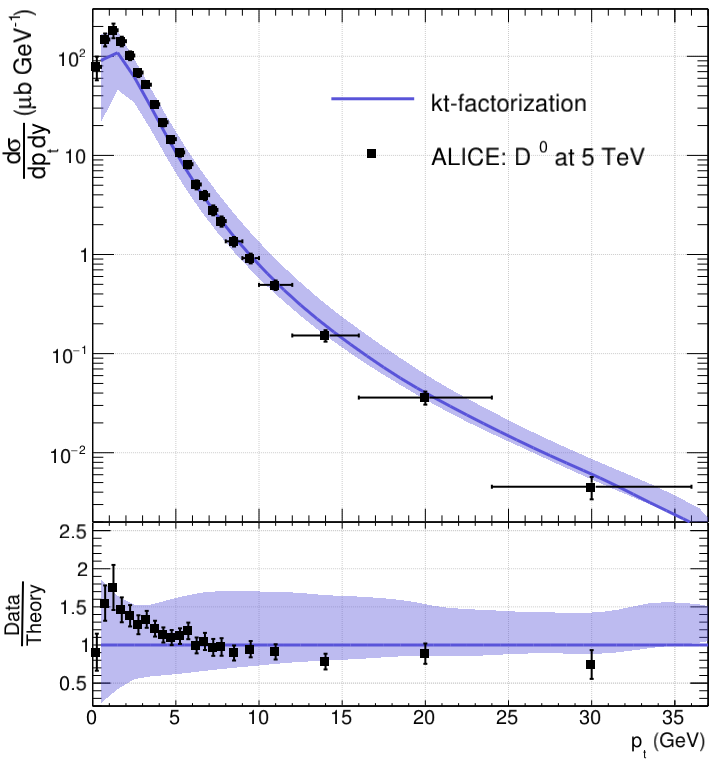}
\caption{$D^0$ production at 5 TeV compared with ALICE data \cite{al5D2021}. \label{d05}}
\end{figure}

\begin{figure}[!h]
\centering
 \includegraphics[width=22pc]{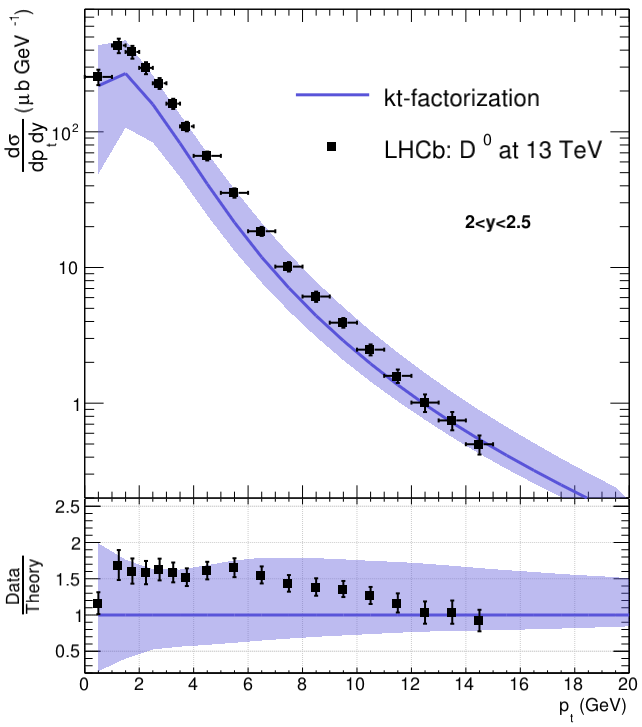}
\caption{$D^0$ production at 13 TeV compared with LHCb data \cite{lhcb13d} in the rapidity range $2<y<2.5$. \label{d13lhcb}}
\end{figure}

Similar results obtained at 7 TeV for $D^{*^+}$ and $D^+_s$ are presented at the end of this paper in Figs. \ref{dpss1} and \ref{dpss2}. We turn now our attention to ALICE measurement at 5 TeV \cite{al5D2021}. The $D^0$ transverse momentum distribution is presented in Fig. \ref{d05}.
Within uncertainties, the agreement with our calculations is excellent. We can also explore the production of $D$ mesons at larger rapidities and energies. In Fig.~\ref{d13lhcb}, we compare our calculations with LHCb data \cite{lhcb13d} at 13 TeV in the rapidity range $2<y<2.5$.
Using the same set of parameters, we obtained a description of experimental data of similar quality compared to the central rapidity case. What has changed is the relative contributions of flavor excitation and creation processes.
\begin{figure}[!h]
\centering
 \includegraphics[width=26pc]{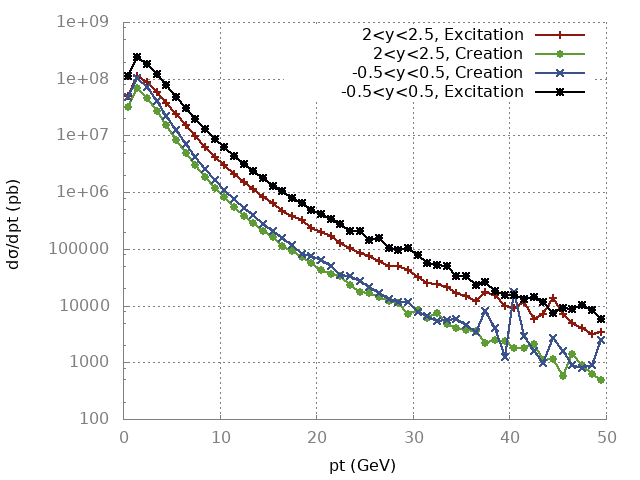}
\caption{Comparison of flavor-excitation and flavor-creation cross sections at central rapidity (7 TeV) and forward rapidity (13 TeV). The green and red curves are closer than the blue and black curves. \label{compCharm}}
\end{figure}
Indeed, in Fig.~\ref{compCharm}, we observe that at forward rapidity the two contributions are closer (green and red curves). The interplay between flavor-excitation and flavor-creation processes being absent in an FFNS at order $\mathcal{O}(\alpha_s^2)$ is probably another good reason for using a GM-VFNS.\\

\begin{figure}[!h]
\centering
 \includegraphics[width=24pc]{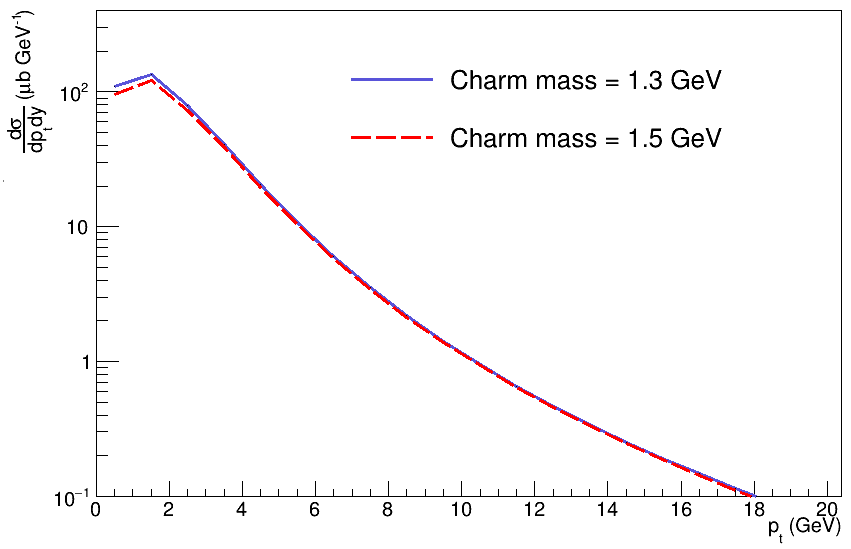}
\caption{$D^0$ transverse momentum distribution for two different charm masses. \label{dpmass}}
\end{figure}

Finally, we quickly discuss other theoretical uncertainties related to our calculations. In Fig.~\ref{dpmass}, we show the result of varying the charm mass from $1.3$ to $1.5$ GeV.
We observe only a small effect at small transverse momentum. Similarly, varying the $p_t$ cuts described in section \ref{seckatie} by a factor of 2 has only a small impact. The uncertainty related to the choice for the fragmentation function has been studied in \cite{MaSz}. Note that we use the same fragmentation function with the same value of the parameter $\epsilon_c$. We did not evaluate the uncertainty related to the choice of the hard scale discussed in \cite{gui1}, as it is not conventional.\\

\subsection{$B$ meson production}
Changing $Q$ by $b$ in (\ref{fcre}) and (\ref{fexc}) gives the complete list of the processes considered. In particular, we did not take into account the $cb\to cb$ process. In agreement with the CT14 PDFs, the bottom mass is set to $m_b=4.75$ GeV. We use the Peterson fragmentation function with $\epsilon_b=0.01$ and the factorization scale Eq.~(\ref{scalekt}) with $m_{t,i}=\sqrt{p_{t,i}^2+m_b^2}$. In agreement with the discussion in \cite{fonll2}, we choose the fragmentation fraction $f(b\to B^+)=$ $f(b\to B^0)=0.403$.\\

\begin{figure}[!h]
\centering
 \includegraphics[width=22.5pc]{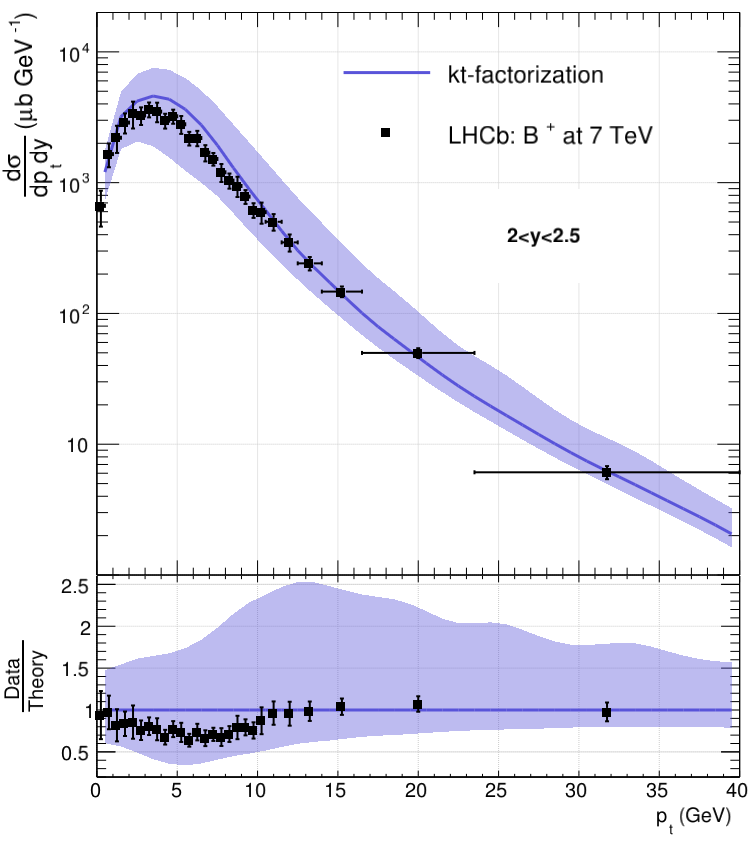}
\caption{$(B^+ + B^-)$ production at 7 TeV compared with LHCb data \cite{lhcbB}. \label{B7225}}
\end{figure}

\begin{figure}[!h]
\centering
 \includegraphics[width=22pc]{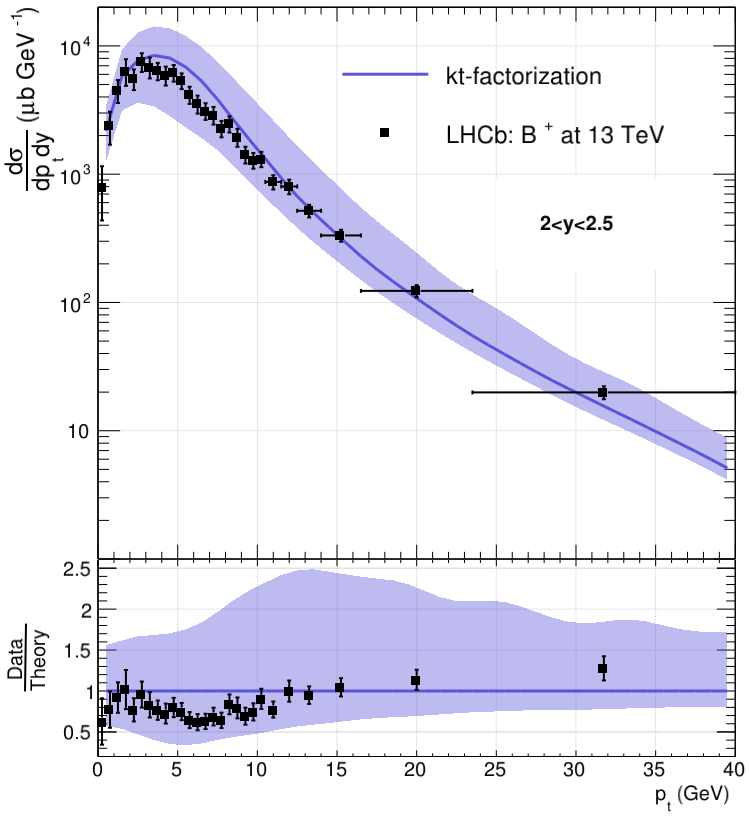}
\caption{$(B^+ + B^-)$ production at 13 TeV compared with LHCb data \cite{lhcbB}. \label{B13225}}
\end{figure}

The LHCb Collaboration measured the $B^\pm$ double-differential cross sections at 7 and 13 TeV, in the rapidity range $2<y<4.5$ \cite{lhcbB}. In Fig.~\ref{B7225}, we compare our results to LHCb data at 7 TeV and $2<y<2.5$.
We observe a global good agreement with our central predictions, with a slight overestimation at $p_t\sim 5$ GeV. All experimental data lay within theoretical uncertainties, estimated by varying the factorization scale. The uncertainty band is broader than for $D$ mesons, which is due to the dependence of the bottom uPDF with $\mu$, and to the fact that the main contribution is given by the flavor-excitation process $gb\to gb$. Changing the Peterson for the  Kartvelishvili et al. fragmentation function \cite{KaLiPe}
\begin{equation}
    D(z)=(\alpha+1)(\alpha+2)z^\alpha (1-z),
\end{equation}
with $\alpha=7$ gives a similar result. Our calculations at 13 TeV, presented in Fig.~\ref{B13225}, show a similar agreement with LHCb data.

\subsection{$B$-meson production at very large rapidity}
\begin{figure}[!h]
\centering
 \includegraphics[width=22pc]{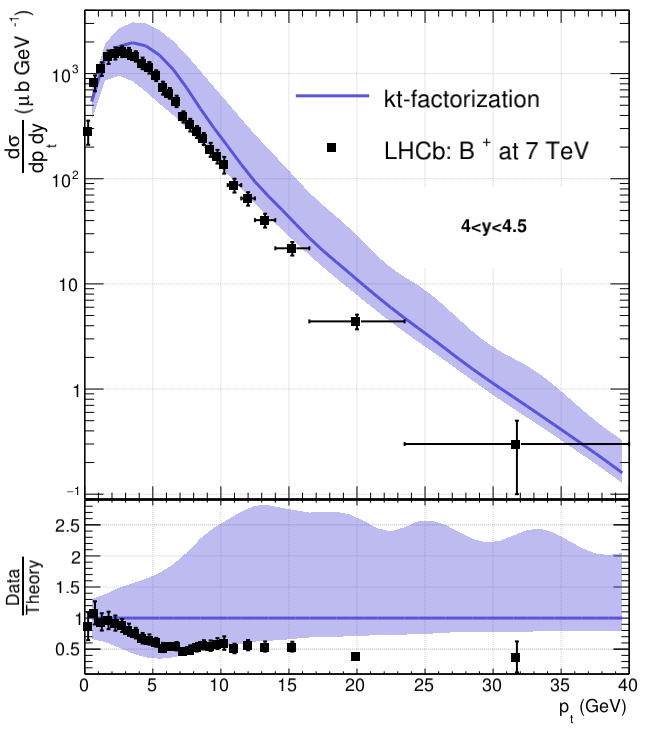}
\caption{$(B^+ + B^-)$ production at 7 TeV compared with LHCb data \cite{lhcbB} in the rapidity range $4<y<4.5$. \label{largeylhcb}}
\end{figure}

Keeping the same fragmentation function, we observe a deviation between our calculations and LHCb data in the rapidity range $4<y<4.5$, see Fig. \ref{largeylhcb}.
Changing parameter $\epsilon_b$ of the fragmentation function restores the agreement between theory and experiment. However, this change is unwanted as it implies a rapidity-dependent fragmentation function which is not part of the $k_t$-factorization formalism, at least in its simplest formulation.

Our understanding of Fig. \ref{largeylhcb} is that it shows the limit of uPDFs built from collinear PDFs. 
\begin{figure}[!h]
\centering
 \includegraphics[width=19pc]{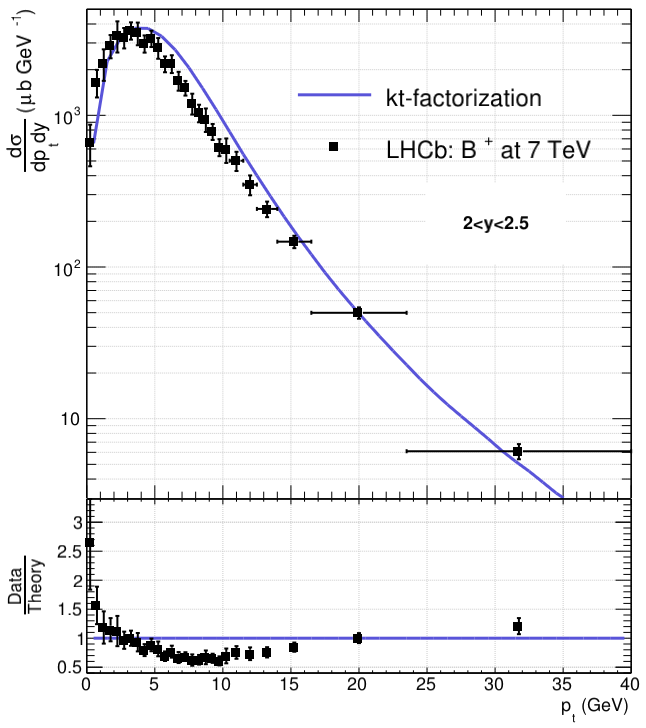}
 \includegraphics[width=19pc]{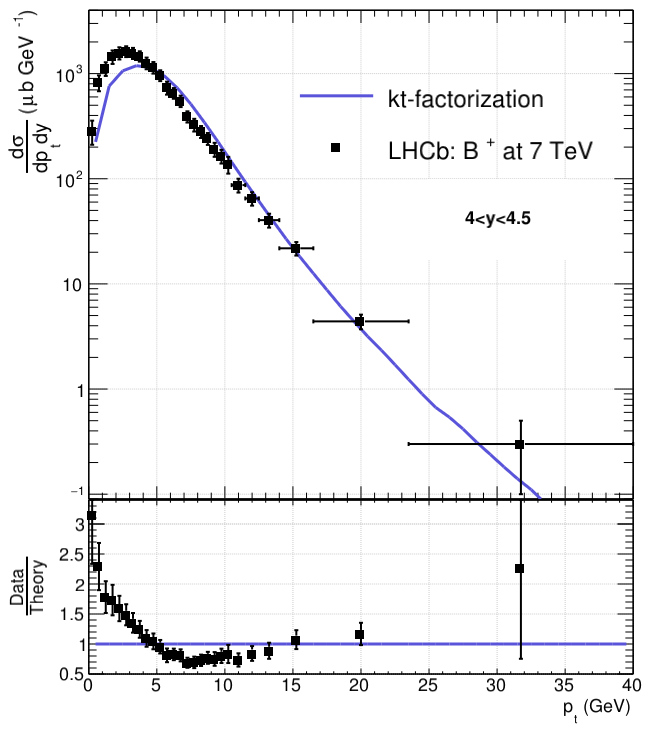}
\caption{Results obtained with the PB uPDFs \cite{PB}. The LHCb data \cite{lhcbB} are described satisfactorily in both rapidity ranges $2<y<2.5$ and $4<y<4.5$. \label{PBlhcb}}
\end{figure}
Indeed, observables at large rapidities trigger smaller and larger values of $x$ compared to central rapidity. From Figs. \ref{unpdf1} and \ref{unpdf2}, we observe that the agreement between integrated uPDFs and collinear PDFs at $x=0.01$ is not as good as at $x=0.001$, and the situation is even worse at larger $x$. The agreement between integrated uPDFs (built from collinear PDFs) and collinear PDFs is, in general, not perfect because relation~(\ref{undens}) holds only approximately. This relation presents other issues discussed in Ref. \cite{gui3}. We expect that uPDFs extracted directly from data will give a better description, and it seems to be the case. In Fig \ref{PBlhcb}, we show a comparison between LHCb data for $B$ mesons and our calculations, with our uPDFs replaced by the PB uPDFs \cite{PB}. The latter are obtained in a GM-VFNS by fitting experimental data. Note that the use of the PB uPDFs, extracted at the NNLO, with our NLO off-shell cross section is not consistent. However, in a VFNS, the numerical impact of this inconsistency is small. With the PB uPDFs, the agreement between theory and experiment is good, including in the rapidity range $4<y<4.5$.\footnote{Here, we concentrate on the region $p_t>$ few GeV. The underestimation at low $p_t$ is expected as the PB uPDFs have been built in a VFNS at order $\mathcal{O}(\alpha_s^3)$. From the discussion of Sec.~\ref{seckt}, it is clear that $\text{PDF}_{\text{NNLO}}^{\text{VNFS}}<\text{PDF}_{\text{NLO}}^{\text{VNFS}}$, but the effect is more visible at low $p_t$.}

\newpage

\section{Comparison with theory \label{seccompth}}
\begin{figure}[!h]
\centering
 \includegraphics[width=22pc]{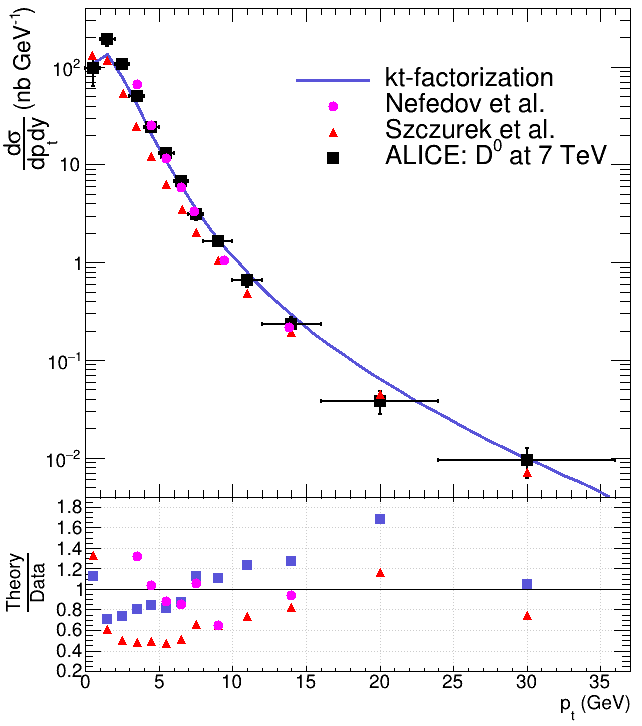}
\caption{Comparison of Nefedov et al. \cite{KaNeSaC} and Szczurek et al. results (extracted from ALICE \cite{al7D2017}) with our calculations and ALICE data \cite{al7D2017}.\label{comp1}}
\end{figure}

In this section, we discuss other $k_t$-factorization calculations and perform a comparison with our results. In Fig.~\ref{comp1} we present ALICE data for $D^0$ mesons at 7 TeV along with the results obtained by Nefedov et al. and Szczurek et al.
These theoretical results have been extracted from Refs. \cite{al7D2017} and \cite{KaNeSaC}. The calculations by Nefedov et al. include the $R+R\to g$ and $R+R\to c +\bar{c}$ processes, where $R$ is a reggeized gluon. The outgoing gluon is transformed into a $D$ meson by a scale-dependent fragmentation function $D_{g\to D}(z,\mu^2)$. We observe a good agreement of Nefedov et al. calculations with data but on a limited $p_t$ range. It should be mentioned that the description of low-$p_t$ data is the hardest part. Szczurek et al. calculations include the $g+g\to c+\bar{c}$ process and do a fair description of the data. However, the situation has been improved by our calculations, which show a better accuracy and stability.

\begin{figure}[!h]
\centering
 \includegraphics[width=22pc]{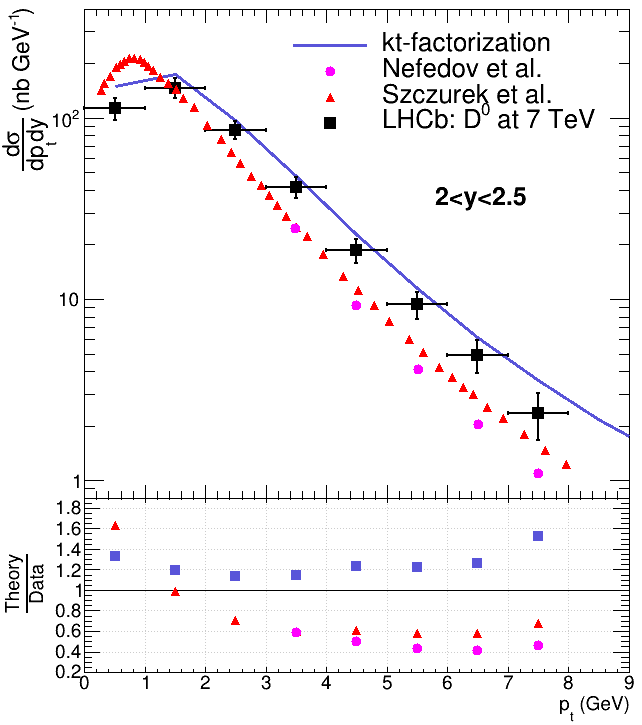}
\caption{Comparison of Nefedov et al. \cite{KaNeSaC} and Szczurek et al. results \cite{MaSz} with our calculations and LHCb data \cite{lhcb7d} at 7 TeV and $2<y<2.5$. \label{comp2}}
\end{figure}

The situation is quite different at forward rapidity, where we compare Nefedov et al. \cite{KaNeSaC} and Szczurek et al. \cite{MaSz} results with LHCb data \cite{lhcb7d}, see Fig. \ref{comp2}. Indeed, the results obtained by these two groups are less satisfying, while our calculations show the same accuracy.
They use the WMR approach with the angular ordering cutoff to build uPDFs from collinear PDFs. The main point is that they choose a set of PDFs determined in a VFN scheme. In section \ref{seckt} we reminded that once the order and scheme of PDFs have been fixed, there is no choice left on the partonic processes to be included. Considering their choice of PDFs, we believe that Szczurek et al. would improve their computation of the partonic cross section by applying the
VFN scheme, and therefore include the non-negligible $c+g\to c+g$ process. The absence of this process explains the observed underestimation. In fact, the underestimation should be even worse, but it is partially compensated by the use of the too large angular-ordered WMR uPDFs \cite{gui2,gui3}.\footnote{The fact that the angular-ordered WMR uPDFs are too large has also been mentioned in \cite{HaKeLe}, figure 6.} The solution proposed in \cite{Maciula:2019izq} is not in agreement with factorization, since the authors increase the $D$-meson cross section by adding some $\mathcal{O}(\alpha_s^3)$ and $\mathcal{O}(\alpha_s^4)$ corrections to the partonic cross section, but keep working with $\mathcal{O}(\alpha_s^2)$ PDFs (for default calculations). A similar discussion could apply to Nefedov et al. calculations since they use the same uPDFs, but the situation is unclear as they include a $2 \to 1$ process.\footnote{We should also mention that due to the large $k_t$ tail of the angular-ordered uPDFs, the upper bound of integration in Eq. (\ref{ktfac}) does matter. This information is in general not available.} 

\begin{figure}[!h]
\centering
 \includegraphics[width=22pc]{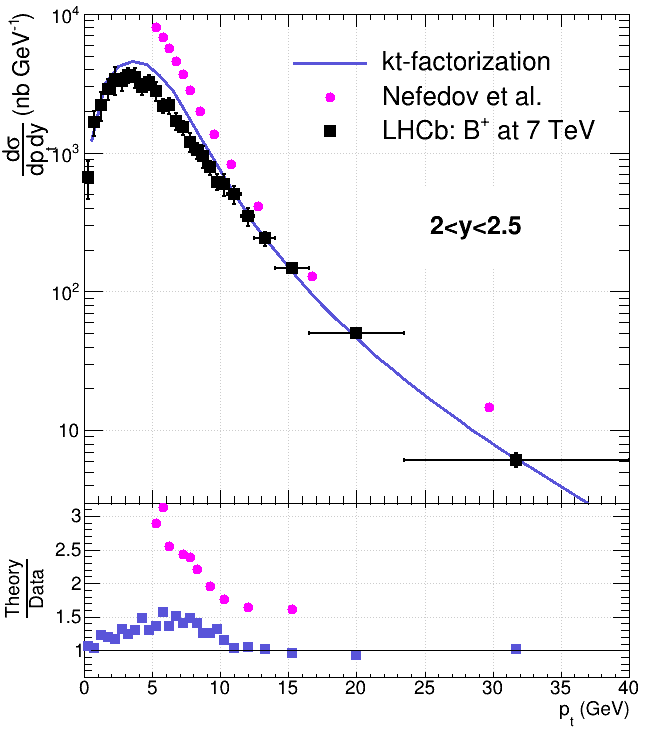}
\caption{Comparison of Nefedov et al.'s calculations \cite{KaNeSaB} with our calculations and LHCb data \cite{lhcbB} at 7 TeV and $2<y<2.5$. We performed a linear interpolation of theory data points in order to compute the ratio to experiment. The ratio between Nefedov et al. and LHCb data at $p_t = 20$ GeV is not shown because the distance between the two last pink circles is too large and linear interpolation not accurate enough. \label{comp3}}
\end{figure}

\begin{figure}[!h]
\centering
 \includegraphics[width=22pc]{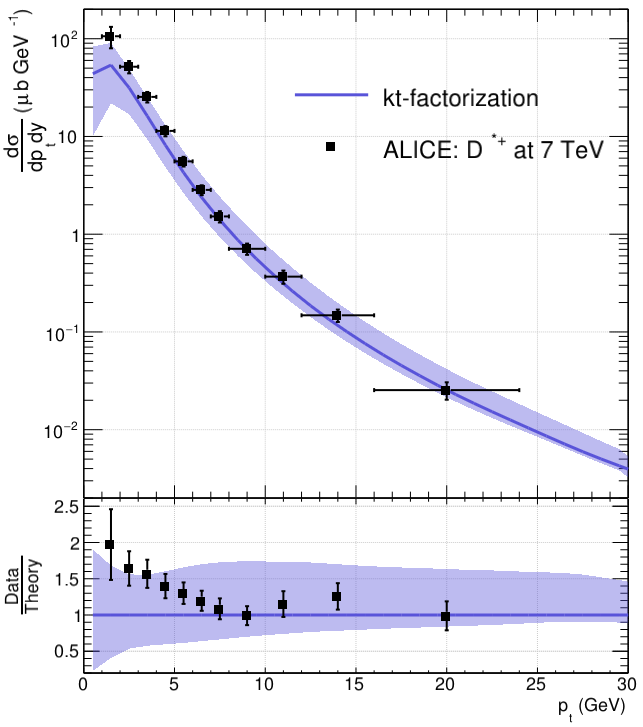}
\caption{Results for $D^{*^+}$ compared to ALICE data at 7 TeV \cite{al7D2017}. \label{dpss1}}
\end{figure}

\begin{figure}[!h]
\centering
 \includegraphics[width=22pc]{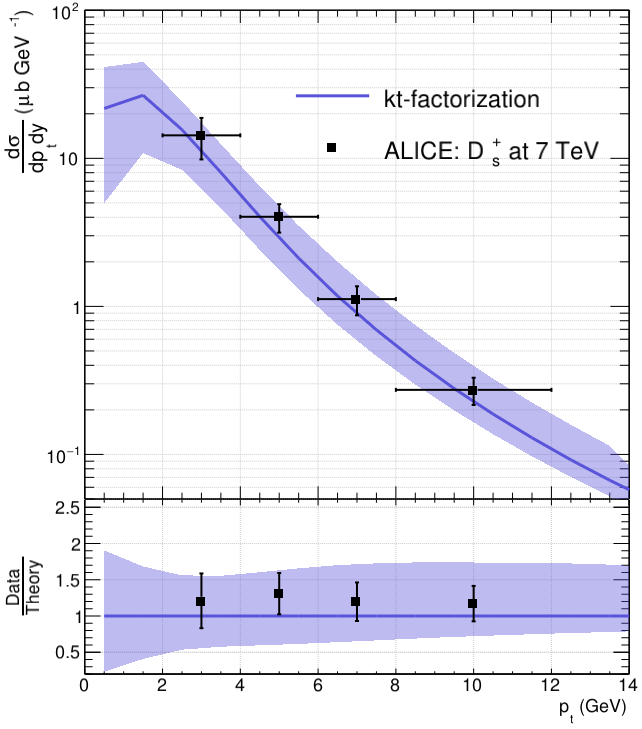}
\caption{Results for $D^+_s$ compared to ALICE data at 7 TeV \cite{al7D2017}. \label{dpss2}}
\end{figure}

Finally, the comparison of LHCb data for $B^+$ mesons with calculations by Nefedov et al. \cite{KaNeSaB} is presented in Fig.~\ref{comp3}. 
Note that they obtained better results at $3<y<3.5$, with only a reasonable  overestimation of the experimental cross section at $p_t>15$. In conclusion, $k_t$-factorization results have been improved by our $\mathcal{O}(\alpha_s^2)$ GM-VFNS calculations. We believe this is directly related to the consistent use of factorization. Insisting on the importance of working fully either in a VFNS or in a FFNS was the main goal of this work. Our results can be improved further by including the $2\to 1$ process, discussed quickly in section \ref{sec12}, and scale-dependent fragmentation functions.

\section{Treatment of the $2\to 1$ process \label{sec12}}
It is sometimes believed that there is a double counting if we take into account both the $gg\to Q\bar{Q}$ and $gQ\to gQ$ processes. It is not the case since double counting happens between two different orders, for instance, when one adds the NLO contribution to a LO calculation. There is consequently no double counting between two LO or NLO processes. On the opposite, the leading-order contribution in $k_t$-factorization is given by the $2\to 1$ process, and we can expect double counting between this process and the $2\to 2$ processes used in this work. The goal of this section is to discuss this in more detail. \\

It is interesting to note that the LO and NLO graphs for deep inelastic scattering are similar to the $2\to 1$ and $2\to 2$ processes in $k_t$-factorization. At leading order, the virtual photon scatters from a (heavy) quark producing an on-shell quark\footnote{The two differences with the $2\to 1$ process in $k_t$-factorization is that the photon should be replaced by a gluon and the initial quark can be off-shell.}. At next-to-leading order, we find $2\to 2$ processes such as $\gamma^*+g\to Q+\bar{Q}$. As explained in \cite{acot1}, there is a double counting between the LO and NLO contributions, because the production of the second particle in the NLO process is also accounted for by the evolution equation of the LO process. A subtraction term is required, and the cross section reads
\begin{equation}
\sigma=\sigma^{\text{LO}}+\sigma^{\text{NLO}}-\text{subtraction term}.\label{fullcs}
\end{equation}
In principle, finding the subtraction term is automatic. One should apply first collinear factorization at the partonic level, allowing one to determine $\hat{\sigma}$, where the hat means that this partonic cross section is free of infrared divergences. In DIS at order $\alpha_s$, the partonic cross section for heavy-quark production with the process initiated by the parton of flavor $a$ is
\begin{equation}
\sigma_a^{(1)}=\sum_b (f_a^{b(0)} \otimes \hat{\sigma}_b^{(1)}+f_a^{b(1)}\otimes\hat{\sigma}_b^{(0)}),
\end{equation}
where $(0)$ and $(1)$ refer to factors proportional to $\alpha_s^0$ and $\alpha_s^1$, respectively. The functions $f_a^{b(n)}$ can be computed perturbatively, and correspond to the distribution of the parton $b$ inside the parton $a$. The cross section $\hat{\sigma}_b^{(n)}$ is associated to the partonic process $\gamma^*+b\to Q+X$. Ref. \cite{acot1} gives an explicit example for $a=g$. Using the fact that $\hat{\sigma}_a^{(0)}=\hat\sigma_a^{(0)}$ and $f_a^{b(0)}(\xi)=\delta_a^b\delta(1-\xi)$, we have
\begin{equation}
\hat{\sigma}_a^{(1)}=\sigma_a^{(1)}-\sum_b f_a^{b(1)}\otimes\hat{\sigma}_b^{(0)}=\sigma_a^{(1)}- f_a^{Q(1)}\otimes\hat{\sigma}_Q^{(0)},
\end{equation}
where we replaced $b$ by $Q$ in the last equality, the LO process being $\gamma^*+Q\to Q$. This procedure automatically generates a term with a minus sign: the subtraction term. $\hat{\sigma}_a^{(1)}$ is free of infrared divergences due to the cancellation between $\sigma_a^{(1)}$ and $f_a^{b(1)}$. In a second step, factorization is applied at the hadronic level,
\begin{align}
\sigma&= f_{b/h}\otimes \hat{\sigma}_b \nonumber\\
&=f_{Q/h}\otimes \sigma_Q^{(0)}+\sum_a \left(f_{a/h}\otimes \sigma_a^{(1)}-f_{a/h}\otimes f_a^{Q(1)}\otimes\hat{\sigma}_Q^{(0)}\right)
\end{align} 
where $f_{a/h}$ gives the distribution of a parton of flavor $a$ in the hadron $h$. We believe that the fully consistent treatment of heavy-quark production in $k_t$-factorization is given by Eq. (\ref{fullcs}), and we plan to work on this soon. However, it is possible to understand why the $2\to 2$ process alone provides a good description of the experimental data. On the one hand, in DIS, there is a cancellation between the subtraction term and the LO term in the region $\mu>m_Q$ and $\alpha_s \ln(\mu/m_Q)\sim \mathcal{O}(1)$. It is the region studied in this work, and we expect a similar situation in $k_t$-factorization. On the other hand, we can argue that the $2\to 1$ process at NLO, with a loop in the conjugate amplitude, is negligible with respect to the $2\to 2 $ process. Indeed, both contributions are of order $\mathcal{O}(\alpha_s^2)$, but the latter is enhanced by the divergences discussed in Sec.~\ref{seckatie}. Consequently, we can expect that the full contribution, Eq.~(\ref{fullcs}), will be numerically close to the result given by the $2\to 2$ process alone.

\section{Conclusion \label{seccon}}
We have presented for the first time $k_t$-factorization calculations for heavy-quark production, using a GM-VFN scheme. We use uPDFs and an off-shell cross section defined in this scheme at the order $\mathcal{O}(\alpha_s^2)$, sometimes misleadingly called leading order. Indeed, the leading order corresponds to the $2\to 1$ process, and Eq.~(\ref{fullcs}) shows how the LO and NLO contributions should be put together. We expect a nearly complete cancellation between the LO contribution and the subtraction term at moderate $p_t$, as well as a negligible role of the $2\to 1$ process at one loop, justifying the use of the $2\to 2$ contributions alone. However, the implementation of Eq.~(\ref{fullcs}) could be important in the region $\mu \gg m_Q$, with potential application to jet physics.\\

The agreement between our calculations and experimental data is excellent, for both $D$ and $B$ mesons. In particular, the agreement has been improved compared to older $k_t$-factorization calculations, performed either in an FFN scheme or in a mix of VFN and FFN schemes. However, we have seen that at very large rapidity our calculations fail if we use uPDFs built from collinear PDFs. In conclusion, accurate and consistent $k_t$-factorization calculations should be performed following the example of collinear factorization: the uPDFs should be exctrated from data, and the cross section computed in a scheme identical to the one used for the uPDFs.

\section*{Acknowledgments}
BG acknowledges support from Chilean FONDECYT Iniciaci\'on grant 11181126. BG is supported by ANID PIA/APOYO AFB180002 (Chile). AvH is supported by grant no. 2019/35/B/ST2/03531 of the Polish National Science Centre.




\begin{thebibliography}{9}
\bibitem{Cacc} M. Cacciari, Rise and Fall of the Bottom Quark Production Excess, \href{https://arxiv.org/pdf/hep-ph/0407187v1.pdf}{arXiv:hep-ph/0407187 (2004)}.
\bibitem{CaCiHa} S.~Catani, M.~Ciafaloni and F.~Hautmann, Gluon contributions to small x heavy flavour production, \href{https://www.sciencedirect.com/science/article/abs/pii/0370269390916017?via%3Dihub}{Phys. Lett. B 242 (1990) 97-102}.
\bibitem{ktFac} J.C. Collins and R.K. Ellis, Heavy-quark production in very high energy hadron collision, \href{https://www.sciencedirect.com/science/article/abs/pii/0550321391902889}{Nucl. Phys. B 360 (1991) 3-30}.
\bibitem{ktFac2} S. Catani, M. Ciafaloni and F. Hautmann, High energy factorization and small-$x$ heavy flavour production, \href{https://www.sciencedirect.com/science/article/abs/pii/0550321391900553}{Nucl. Phys. B 366(1991) 135-188}.
\bibitem{semihard2} E. M. Levin, M. G. Ryskin, Y. M. Shabelski, A. G. Shuvaev, Heavy quark production in semihard nucleon interaction, Sov. J. Nucl. Phys. 53 (1991) 657.
\bibitem{LiSaZo} A. V. Lipatov, V. A. Saleev, and N. P. Zotov, Heavy Quark Production at the TEVATRON in the Semihard QCD Approach  and  the  Unintegrated  Gluon  Distribution, \href{https://arxiv.org/abs/hep-ph/0112114v3}{arXiv:hep-ph/0112114v3 (2001)}.
\bibitem{jun02} H. Jung, Heavy quark production at the TEVATRON and HERA using $k_t$ factorization with CCFM evolution, \href{https://journals.aps.org/prd/abstract/10.1103/PhysRevD.65.034015}{Phys. Rev. D 65 (2002) 034015}.
\bibitem{JuKrLi} H. Jung, M. Kraemer, A.V. Lipatov, and N.P. Zotov, Heavy Flavour Production at Tevatron and Parton Shower Effects, \href{https://link.springer.com/article/10.1007/JHEP01(2011)085}{JHEP 01 (2011) 085}.
\bibitem{MaSz} R. Maciuła, A. Szczurek, Open charm production at the LHC - $k_t$-factorization approach, \href{https://journals.aps.org/prd/abstract/10.1103/PhysRevD.87.094022}{Phys. Rev. D 87 (2013) 094022}.
\bibitem{ShShSu} Yu. M. Shabelski, A. G. Shuvaev, and I. V. Surnin, Heavy quark production in kt factorization approach at LHC energies, \href{https://www.worldscientific.com/doi/abs/10.1142/S0217751X18500033}{Int. J. Mod. Phys. A33, 1850003 (2018)}.
\bibitem{MaSz2} R. Maciuła, A. Szczurek, Consistent treatment of charm production in higher-orders at tree-level within $k_t$-factorization approach, \href{https://journals.aps.org/prd/abstract/10.1103/PhysRevD.100.054001}{    Phys.Rev.D 100 (2019) 5, 054001}.
\bibitem{gui2} B.~Guiot, Heavy-quark production with $k_t$-factorization: The importance of the sea-quark distribution, \href{https://journals.aps.org/prd/abstract/10.1103/PhysRevD.99.074006}{Phys.Rev.D 99 (2019) 7, 074006}.
\bibitem{gui3} B.~Guiot, Pathologies of the Kimber-Martin-Ryskin prescriptions for unintegrated PDFs: Which prescription should be preferred?, \href{https://journals.aps.org/prd/abstract/10.1103/PhysRevD.101.054006}{Phys.Rev.D 101 (2020) 5, 054006}.
\bibitem{Kniehl:2008qb}
B.~A.~Kniehl, A.~V.~Shipilova and V.~A.~Saleev,
Open charm production at high energies and the quark Reggeization hypothesis,
\href{https://journals.aps.org/prd/abstract/10.1103/PhysRevD.79.034007}{Phys. Rev. D \textbf{79}, 034007 (2009)}.
\bibitem{PB} A. B. Martinez, P. Connor, F. Hautmann, H. Jung, A. Lelek,V. Radescu, and R. Zlebcik, Collinear and TMD parton densities from fits to precision DIS measurements in the parton branching method, \href{https://journals.aps.org/prd/abstract/10.1103/PhysRevD.99.074008}{Phys.Rev.D 99 (2019) 7, 074008}.
\bibitem{tmdlib} F. Hautmann, H. Jung, M. Krämer, P. J. Mulders, E. R. Nocera,
T. C. Rogers and A. Signori, TMDlib and TMDplotter: library and plotting tools for transverse-momentum-dependent parton distributions, \href{https://link.springer.com/article/10.1140/epjc/s10052-014-3220-9}{Eur. Phys. J. C 74 (2014) 3220}.
\bibitem{tmdlib2} N.~A.~Abdulov \textit{et al.}, TMDlib2 and TMDplotter: a platform for 3D hadron structure studies, \href{https://arxiv.org/abs/2103.09741}{arXiv:2103.09741}.
\bibitem{wmr} G. Watt, A.D. Martin and M.G. Ryskin, Unintegrated parton distributions and inclusive jet production at HERA, \href{https://link.springer.com/article/10.1140/epjc/s2003-01320-4}{Eur. Phys. J. C 31(2003) 73}.
\bibitem{ct14} S. Dulat et. al., New parton distribution functions from a global analysis of quantum chromodynamics, \href{https://journals.aps.org/prd/abstract/10.1103/PhysRevD.93.033006}{Phys. Rev. D 93 (2016) no.3, 033006}.
\bibitem{acot1}M. A. G. Aivazis, J. C. Collins, F. I. Olness and W.-K. Tung, Leptoproduction of heavy quarks. II. A unified QCD formulation of charged and neutral current processes from fixed-target to collider energies, \href{https://journals.aps.org/prd/abstract/10.1103/PhysRevD.50.3102}{Phys. Rev. D 50, 3102 (1994)}.
\bibitem{acot2}J. C. Collins, Hard-scattering factorization with heavy quarks: A general treatment, \href{https://journals.aps.org/prd/abstract/10.1103/PhysRevD.58.094002}{Phys. Rev. D 58, 094002 (1998)}.
\bibitem{acot3}W.-K. Tung, S. Kretzer and C. Schmidt, Open heavy flavour production: conceptual framework and implementation issues, J. Phys. G 28, 983 (2002) \href{https://arxiv.org/abs/hep-ph/0110247}{[hep-ph/0110247]}.
\bibitem{vanHameren:2016kkz}
A.~van Hameren, KaTie : For parton-level event generation with $k_T$-dependent
initial states, \href{https://www.sciencedirect.com/science/article/abs/pii/S0010465517303880?via%3Dihub}{Comput. Phys. Commun. \textbf{224} (2018), 371-380}.
\bibitem{Alwall:2006yp} J.~Alwall, \textit{et al.}, A Standard format for Les Houches event files, \href{https://www.sciencedirect.com/science/article/abs/pii/S0010465506004164?via%3Dihub}{Comput. Phys. Commun. \textbf{176} (2007), 300-304}.
\bibitem{vanHameren:2012if} A.~van Hameren, P.~Kotko and K.~Kutak, Helicity amplitudes for high-energy scattering, \href{https://link.springer.com/article/10.1007/JHEP01(2013)078}{JHEP \textbf{01} (2013), 078}.
\bibitem{vanHameren:2013csa} A.~van Hameren, K.~Kutak and T.~Salwa, Scattering amplitudes with off-shell quarks, \href{https://www.sciencedirect.com/science/article/pii/S0370269313008460?via%3Dihub}{Phys. Lett. B \textbf{727} (2013), 226-233}.
\bibitem{pescsc} C. Peterson, D. Schlatter, I. Schmitt, P.M. Zerwas, Scaling violations in inclusive $e^+ e^-$ annihilation spectra, \href{https://journals.aps.org/prd/abstract/10.1103/PhysRevD.27.105}{Phys. Rev. D27 (1983) 105}.
\bibitem{fonll}   M.~Cacciari, M.~Greco and P.~Nason,``The p(T) spectrum in heavy-flavor hadroproduction", \href{https://iopscience.iop.org/article/10.1088/1126-6708/1998/05/007}{JHEP {\bf 05} (1998) 007};\\
M.~Cacciari, S.~Frixione and P.~Nason, ``The p(T) spectrum in heavy-flavor photoproduction”, \href{https://iopscience.iop.org/article/10.1088/1126-6708/2001/03/006}{JHEP {\bf 03} (2001) 006}.
\bibitem{al7D2017}S. Acharya et al., Measurement of D-meson production at mid-rapidity in pp collisions at $\sqrt{s}=7$ TeV, \href{https://link.springer.com/article/10.1140/epjc/s10052-017-5090-4}{Eur.Phys.J.C 77 (2017) 8, 550}.
\bibitem{KaNeSaC} A.~V.~Karpishkov, M.~A.~Nefedov, V.~A.~Saleev and A.~V.~Shipilova, Open charm production in the parton Reggeization approach: Tevatron and the LHC, \href{https://journals.aps.org/prd/abstract/10.1103/PhysRevD.91.054009}{Phys. Rev. D 91 (2015) 5, 054009}.
\bibitem{KaNeSaB} A.~V.~Karpishkov, M.~A.~Nefedov, V.~A.~Saleev and A.~V.~Shipilova, B-meson production in the Parton Reggeization Approach at Tevatron and the LHC, \href{https://www.worldscientific.com/doi/abs/10.1142/S0217751X15500232}{Int. J. Mod. Phys. A 30 (2015) 04n05, 1550023}.
\bibitem{al5D2021}S. Acharya et al., Measurement of beauty and charm production in pp collisions at $\sqrt{s}=5$ TeV, via non-prompt and prompt D mesons, \href{https://link.springer.com/article/10.1007/JHEP05(2021)220}{JHEP 05 (2021) 220}.
\bibitem{lhcb13d}LHCb Collaboration, Roel Aaij et al., Measurements of prompt charm production cross-sections in $pp$ collisions at $\sqrt{s}=13$ TeV, \href{https://link.springer.com/article/10.1007/JHEP03(2016)159}{JHEP 03 (2016) 159}, \href{https://link.springer.com/article/10.1007%2FJHEP09%282016%29013}{JHEP 09 (2016) 013} (erratum),  \href{https://link.springer.com/article/10.1007%2FJHEP05%282017%29074}{JHEP 05 (2017) 074} (erratum)
\bibitem{gui1} Benjamin Guiot, Hard scale uncertainty in collinear factorization: Perspective from kt -factorization, \href{https://journals.aps.org/prd/abstract/10.1103/PhysRevD.98.014036}{Phys.Rev.D 98 (2018) 1, 014036}.
\bibitem{fonll2}M. Cacciari et al., Theoretical predictions for charm and bottom production at the LHC, \href{https://link.springer.com/article/10.1007/JHEP10(2012)137}{JHEP 10 (2012) 137}.
\bibitem{lhcbB} R. Aaij et al., Measurement of the $B^{\pm}$ production cross-section in pp collisions at $\sqrt{s}=7$ and 13 TeV, \href{https://link.springer.com/article/10.1007%2FJHEP12%282017%29026}{JHEP 12 (2017) 026}.
\bibitem{KaLiPe} V. G. Kartvelishvili, A. K. Likhoded, V. A. Petrov, \href{https://www.sciencedirect.com/science/article/abs/pii/0370269378906536?via%3Dihub}{Phys.Lett.B 78 (1978) 615-617}.
\bibitem{lhcb7d} LHCb Collaboration, R. Aaij et al.,Prompt charm production in pp collisions at sqrt(s)=7 TeV, \href{https://www.sciencedirect.com/science/article/pii/S0550321313000965?via%3Dihub}{Nucl.Phys. B871, 1-20 (2013)}.
\bibitem{HaKeLe} F. Hautmann, L. Keersmaekers, A. Lelek, A.M. Van Kampen, Dynamical resolution scale in transverse momentum distributions at the LHC, \href{https://www.sciencedirect.com/science/article/pii/S0550321319302810?via%3Dihub}{Nucl. Phys. B 949 (2019) 114795}.
\bibitem{Maciula:2019izq} R.~Maciuła and A.~Szczurek, Consistent treatment of charm production in higher-orders at tree-level within $k_T$-factorization approach, \href{https://journals.aps.org/prd/abstract/10.1103/PhysRevD.100.054001}{Phys. Rev. D 100, no.5, 054001 (2019)}.
\end{thebibliography}
\end{document}